\documentclass[12pt]{iopart}
\usepackage{graphicx}

\begin{document}

\title[]{The contribution of spin torque to spin Hall coefficient and spin motive force
in spin-orbit coupling system}

\author{Yong-Ping Fu$^{1,2}$, Dong Wang$^{1,2}$, F. J. Huang$^{1,2}$, Y. D. Li$^{2}$, W. M. Liu$^{1}$}

\address{$^1$Beijing National Laboratory for Condensed Matter Physics,
Institute of Physics, Chinese Academy of Sciences, Beijing 100080,
China}
\address{$^2$School of Physics, Yunnan's University, Kunming 650091,
China}
\date{\today }

\ead{ynufyp@sina.cn; fuyongping44@yahoo.com.cn}
\begin{abstract}
We derive rigorously the relativistic angular momentum conservation
equation by means of quantum electrodynamics. The novel
nonrelativistic spin current and torque in the spin-orbit coupling
system, up to the order of $1/c^{4}$, are exactly investigated by
using Foldy-Wouthuysen transformation. We find a perfect spin Hall
coefficient including the contribution of spin torque dipole. A
novel spin motive force, analogue to the Lorentz force, is also
obtained for understanding of the spin Hall effect.
\end{abstract}


\maketitle
\section{Introduction}
Spintronics has become a fast developing field since it developed.
The transport concerned aspect of the carriers' spin degree and the
spin Hall effect [1-3] were paid a lot of attention recently. In
order to describe the spin transport properly, the definition of
spin current was discussed and various theories of spin current have
been established \cite{1,4}. In a traditional review, the spin
current was presented in terms of an anticommutator of the velocity
and the spin, $(1/2) \varphi^{+}\{\mathbf{v},\mathbf{s}\}\varphi$.
However, under such a definition one of problem is that there is not
a conjugate spin force to link the spin current. Therefore, the
Onsager relation can not be established \cite{29}. Furthermore,
because the spin has its own dynamics in its Hilbert space, the
current with both spin and spatial degree is not conserved due to
the spin-orbit coupling. With the consideration of a spin torque, a
source in the spin continuity equation can be achieved. Previous
investigations in the spin torque depend on the spin relaxation time
[7-12]. To our knowledge, an explicit torque beyond of approximation
of spin relaxation time has not been established yet.

In the studies of spin Hall effect, the experiments and theories
focus on the spin Hall coefficient $\sigma_{SH}$ [4, 5, 13-31]. In
comparison of Ohm's law in electronics responded to the applied
electric field a spin current $j_{s}^{kl}$ is generated,
$j_{s}^{kl}=\sigma_{SH}\varepsilon ^{lkm}E^{m}$ \cite{1}. Recent
studies shew that the spin Hall coefficient $\sigma_{SH}$ not only
includes the contribution of the conventional spin current, but also
the torque dipoles which are contained in semiconductor models with
the effect of disorder \cite{29,34}. However, those contributions
from the torque dipoles have not been clearly found yet.

Based on the above considerations, the consistency of quantum
electrodynamics and Noether's theorem in the derivation of the exact
conservation equation for the relativistic angular momentum was
suggested \cite{31,32}. It is found that the spin current including
a correction is different from the traditional definition. In the
application the spin Hall conductivity $\sigma_{SH}$ involved the
correction can be obtained. Under the requirement of the Onsager
relation the spin force is found to relate to the spin Hall
coefficient, therefore, relate the topological aspect of systems
with the spin-orbit coupling.

\section{Spin continuity equation}
Let us firstly consider the relativistic Lagrangian with Dirac
fields $\Psi$ and $\bar{\Psi}$ coupled to an electromagnetic field
$A^{\mu}$, $\mathcal{L} =\mathcal{L}_{\mathrm{D}}
+\mathcal{L}_{\mathrm{em}} +\mathcal{L}_{\mathrm{int} }$, where
$\mathcal{L}_{\mathrm{D}}=\bar{\Psi}(i\hbar c\gamma^{\mu}
\partial_{\mu}-mc^{2})\Psi$ describes the free Dirac fields of spin $1/2$,
$\mathcal{L}_{\mathrm{em}}= -(1/4)F^{\mu\nu}F_{\mu\nu}$ is the
Lagrangian of electromagnetic field, where $F^{\mu\nu}=
\partial^{\mu}A^{\nu}-\partial^{\nu }A^{\mu}$ , the interaction
between Dirac fields and electromagnetic field is given by
$\mathcal{L}_{\mathrm{int}}=-e\bar{\Psi}\gamma^{\mu}A_{\mu}\Psi$,
and the four-vector $\gamma^{\mu}$ is represented as
$\gamma^{\mu}=(\gamma^{0},\gamma)$ in terms of Pauli matrices
$\sigma$.

The energy-momentum tensor of a gauge invariant form is found to be
$\theta^{\mu\nu}=\theta_{\mathrm{D}}^{\mu\nu}+\theta_{\mathrm{em}}^{\mu\nu
}+\theta_{\mathrm{int}}^{\mu\nu}$, where $\theta_{\mathrm{D}}^{\mu\nu}%
=\bar{\Psi}i\hbar c \gamma^{\mu}\partial^{\nu}\Psi-g^{\mu\nu}\mathcal{L}_{\mathrm{D}}%
$,
$\theta_{\mathrm{em}}^{\mu\nu}=-F^{\mu\sigma}\partial^{\nu}A_{\sigma
}-g^{\mu\nu}\mathcal{L}_{\mathrm{em}}$, and
$\theta_{\mathrm{int}}^{\mu\nu
}=-g^{\mu\nu}\mathcal{L}_{\mathrm{int}}$. Here
$g^{\mu\nu}=g_{\mu\nu}$ is the metric tensor with $g^{00}=1$,
$g^{ii}=-1$ $(i=1,2,3)$ and $g^{\mu\nu}=0$
$(\mu,\nu=0,1,2,3,\mu\neq\nu)$. This energy-momentum tensor
satisfies the conservation law, i.e.,
$\partial_{\mu}\theta^{\mu\nu}=0$. With the tensor the angular
momentum tensor can be written in the form of $M^{\alpha\mu\nu
}=s^{\alpha\mu\nu}+l^{\alpha\mu\nu}$. Here $l^{\alpha\mu\nu}=x^{\mu}%
\theta^{\alpha\nu}-x^{\nu}\theta^{\alpha\mu}$ is the orbital angular
momentum
tensor and $s^{\alpha\mu\nu}=s_{\mathrm{D}}^{\alpha\mu\nu}+s_{\mathrm{em}%
}^{\alpha\mu\nu}$ is spin angular momentum tensor, where $s_{\mathrm{D}%
}^{\alpha\mu\nu}=\left(  \partial\mathcal{L}/\partial\partial_{\alpha}%
\Psi\right)  I_{\mathrm{D}}^{\mu\nu}\Psi$ and
$s_{\mathrm{em}}^{\alpha\mu\nu }=\left(
\partial\mathcal{L}/\partial\partial_{\alpha}A_{\sigma}\right)
(I_{\mathrm{em}}^{\mu\nu})_{\sigma\rho}A^{\rho}$. Considering the
notations $I_{\mathrm{D}}^{\mu\nu}=-i\sigma^{\mu\nu}/2$ and
$(I_{\mathrm{em}}^{\mu\nu
})_{\sigma\rho}=g_{\sigma}^{\mu}g_{\rho}^{\nu}-g_{\rho}^{\mu}g_{\sigma}^{\nu}%
$, it is found $s_{\mathrm{D}}^{\alpha\mu\nu}=i\left(  \hbar
c/4\right)
\bar{\Psi}\gamma^{\alpha}[\gamma^{\mu},\gamma^{\nu}]\Psi$ and $s_{\mathrm{em}%
}^{\alpha\mu\nu}=A^{\mu}F^{\alpha\nu}-A^{\nu}F^{\alpha\mu}$. The
corresponding
conservation law for the total angular momentum is $\partial_{\alpha}%
M^{\alpha\mu\nu}=0$.

\begin{figure}[ptb]
\begin{picture}(0,140)(-130,-90)
\thicklines \put(0,0){\line(3,2){48}} \put(70,0){\line(3,2){48}}
\put(0,0){\line(1,0){70}} \put(48,32){\line(1,0){70}}
\put(0,-7){\line(0,1){7}} \put(70,-7){\line(0,1){7}}
\put(118,25){\line(0,1){7}} \put(0,-7){\line(1,0){70}}
\put(70,-7){\line(3,2){48}} \put(48,25){\circle*{4}}
\put(85,17){\circle*{4}} \put(48,25){\vector(3,2){20}}
\put(48,25){\vector(1,0){18}} \put(48,25){\vector(0,-1){11}}
\put(85,17){\vector(-3,-2){10}} \put(85,17){\vector(1,0){26}}
\put(85,17){\vector(0,1){15}}
\put(38,10){$s^{l}$}\put(40,37){$J^{kl}_{s}\!\!,\!f^{k}$}\put(66,20){$E^{m}$}
\put(48,5){$J^{kl}_{s}\!\!,\!f^{k}$} \put(112,13){$E^{m}$}
\put(85,33){$s^{l}$} \put(160,20){\line(3,2){35}}
\put(160,-70){\line(3,2){35}} \put(160,20){\line(0,-1){90}}
\put(195,43){\line(0,-1){90}} \put(190,15){\circle*{4}}
\put(190,15){\vector(-3,-2){15}} \put(190,15){\vector(0,1){15}}
\put(162,12){$J^{kl}_{s}$}\put(162,-3){$f^{k}$}
\put(185,30){$s^{l}$} \put(165,-40){\vector(3,2){17}}
\put(165,-40){\vector(0,-1){15}} \put(165,-40){\circle*{4}}
\put(180,-23){$J^{kl}_{s}$}\put(180,-40){$f^{k}$}
\put(163,-63){$s^{l}$} \put(167,-10){\vector(1,0){40}}
\put(211,-12){$E^{m}$}\put(150,-10){\line(1,0){10}}
\put(0,-80){\line(3,2){48}} \put(70,-80){\line(3,2){48}}
\put(0,-80){\line(1,0){70}} \put(48,-48){\line(1,0){70}}
\put(0,-87){\line(0,1){7}} \put(70,-87){\line(0,1){7}}
\put(118,-55){\line(0,1){7}} \put(0,-87){\line(1,0){70}}
\put(70,-87){\line(3,2){48}} \put(50,-63){\circle*{4}}
\put(50,-63){\vector(-3,-2){10}} \put(50,-63){\vector(1,0){26}}
\put(50,-63){\vector(0,1){15}}
\put(50,-47){$s^{l}$}\put(15,-76){$J^{kl}_{s}\!\!,\!f^{k}$}\put(74,-60){$E^{m}$}
\put(100,-55){\vector(0,1){25}}\put(100,-74){\line(0,1){7}}
\put(103,-34){$B^{l}$}
%
\put(15,35){a} \put(150,35){c}\put(15,-46){b}
\end{picture}
\caption{The spin current $J^{kl}_{s}$ and the spin motive force
$f^{k}$ via the spin $\mathbf{s}$ and the electric field
$\mathbf{E}$, where $J^{kl}_{s}$ represents the current of the $l$
component $s^{l}$ of the spin along the direction $k$. (a) the spin
current $J^{kl}_{s}$ and the spin motive force $f^{k}$ in the
spin-orbit coupling system without an external magnetic field, where
$E^{m}$, $s^{l}$ and $J^{kl}_{s}$ (or $f^{k}$) satisfy the
right-hand rule; (b) the spin current and the spin motive force in
the spin-orbit coupling system under an external magnetic field
$\mathbf{B}$ along the $l$ direction; (c) the spin current and the
spin motive force in the
two-dimensional Rashba spin-orbit coupling system.}%
\end{figure}
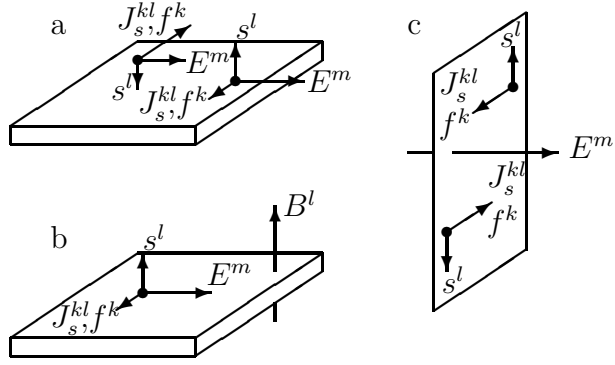

In order to obtain the nonrelativistic form of the conservation law,
the Foldy-Wouthuysen transformation is used in the following
calculations up to $1/c^{4}$. The nonrelativistic wave function is
written in terms of a transformation on the relativistic wave
function $\Psi$, $\Psi^{^{\prime\prime }}=\exp\left[
is^{\prime}\left(  \alpha\right)  \right] \exp\left[ is\left(
\alpha\right)  \right]  \Psi$, where the operators in the
exponential are $is\left( \alpha\right) \equiv\left(
\beta/2mc\right)  \alpha \cdot\ \pi$ and $is^{\prime}\left(
\alpha\right)  \equiv\left(  i\hbar e/4m^{2}c^{3}\right)
\alpha\cdot\mathbf{E}$. Here $\mathbf{E}$ is the electric field
intensity. Correspondingly the wave function is written in the form
as
$\Psi^{^{\prime\prime}}=(\varphi^{\prime\prime},\chi^{\prime\prime})^{\emph{T}}$,
where $\varphi^{\prime\prime}=\left[  1-s\left( \sigma\right)
^{2}/2\beta^{2}\right]  \varphi$ and $\chi ^{\prime\prime}=\left[
is^{\prime}\left( \sigma\right)
-i\left(  E-e\phi\right)  s\left(  \sigma\right)  /2mc^{2}%
\beta-is\left(  \sigma\right)  ^{3}/3\beta^{3}\right] \varphi$.
Introducing a notion $\eta=i\hbar e\sigma\cdot\mathbf{E}%
-(E-e\phi)\sigma\cdot\pi-(\sigma\cdot \pi)^{3}/6m$,
$\chi^{\prime\prime}$ is presented as $\chi ^{\prime\prime}=\left(
\eta/4m^{2}c^{3}\right)  \varphi$. With the help of formula
$e^{is\left(  \alpha\right) }\hat{O}e^{-is\left(
\alpha\right)  }=\hat{O}+[is,\hat{O}]+[is,[is,\hat{O}%
]]/2+\cdots+[is,[is,\cdots,[is,\hat{O}]\cdots]]/n!+\cdots$ and let
$\hat{O}$ be $M^{0ij}$ and $M^{kij}$, the continuity equation for
the nonrelativistic electronic spin can be obtained. The
nonrelatistivic form of angular momentum conservation law reads
\begin{equation}
\frac{\partial}{\partial t}\rho_{s}^{l}+\nabla^{k}j_{s}^{kl}=T^{l},
\end{equation}
where $j_{s}^{kl}=\left(  \hbar/4m\right)  \varphi^{\dag}\left\{ \pi
^{k},\sigma^{l}\right\}  \varphi$ is the traditional spin current,
which represents the current of the $l$ component of the spin along
the direction $k$. Here we have written the wave function
$\varphi^{\prime\prime}$ as $\varphi$ for the convenient. The spin
density $\rho _{s}^{l}$ is obtained as
\begin{eqnarray}
\fl \rho_{s}^{l} =\frac{\hbar }{2}\varphi ^{+}\sigma ^{l}\varphi +
\frac{\hbar }{4m^{2}c^{2}}\varphi ^{+}(\pi ^{l}\sigma\cdot
\pi\!-\!\pi^{2}\sigma ^{l})\varphi \nonumber\\
+\frac{\hbar^{2} e}{8m^{2}c^{3}}\varphi ^{+}(3B^{l} \!-\! \sigma ^{l}%
\sigma\cdot \mathbf{B})\varphi
+\frac{i\hbar}{8m^{3}c^{4}}\varphi ^{+}[(\sigma\times %
\pi)^{l}\eta \!-\! \eta ^{+}(\sigma\times %
\pi)^{l}]\varphi,
\end{eqnarray}
where magnetic field $\mathbf{B}$ is written out evidently. The
first term in Eq. (2) is nothing but a traditional spin density. The
second term can be written as $\left( \hbar/4m^{2}c^{2}\right)
\varphi^{+}\pi \times(\pi \times \sigma)\varphi$, which indicates
its generation from the spin-orbit coupling. The interaction between
the intrinsic magnetic moment and the external magnetic field is
given by the third term. The last term gives a small correction in
the order of $1/c^{4}$.

Now let us analysis the right hand of Eq. (1), named the spin torque
density $T^{l}$. Up to the same order of the nonrelativistic
approximation, it is found
\begin{eqnarray}
\fl T^{l} = \nabla^{k}\{\frac{i\hbar}{2m}\varphi^{+}\sigma^{k}
(\sigma\times\pi)^{l}\varphi\}+\frac{i\hbar e}
{2mc}\varphi^{+}[\sigma^{l}\sigma\cdot\mathbf{B} - B^{l}]\varphi
\nonumber\\
- \frac{\hbar e}{4m^{2}c^{2}}\varphi^{+}\{\hbar
\lbrack\nabla(\mathbf{E}\cdot
\sigma)\times\sigma]^{l}+2\sigma^{l}%
\pi\cdot\mathbf{E}-2\sigma\cdot \pi\ E^{l}\}\varphi\
\nonumber\\
+ \frac{\hbar^{2}e}{4m^{2}c^{2}}\nabla ^{k}
\{\varphi^{+}[\sigma^{k}(\sigma\times\mathbf{E}
)^{l}+(\sigma\times\mathbf{E})^{k}\sigma^{l}]\varphi\}
\nonumber\\
-\frac{1}{32m^{4}c^{4}}\varphi^{+}(\eta^{+}\{\sigma\cdot
\pi,\{\sigma\cdot \pi,(\pi\times \sigma)^{l}\}\}+\{\sigma\cdot
\pi,\!\{\sigma\!\cdot\! \pi,\!(\pi\!\times\sigma)^{l}\}\}\eta
)\varphi\
\nonumber\\
 +\frac{\hbar}{64m^{4}c^{4}}\nabla^{k}[\varphi^{+}(\eta^{+}
\{\sigma\cdot\pi,\{\sigma\cdot
\pi,\sigma^{k}\sigma^{l}\}\}+\{\sigma\cdot
\pi,\{\sigma\cdot\pi,\sigma^{k} \sigma^{l}\}\}\eta)\varphi].
\end{eqnarray}
Besides of the relativistic correction up to the order of $1/c^{4}$,
the contributions from the spin-orbit coupling and its
nonrelativistic correction are presented by the first and the forth
terms. The second term corresponds to the interaction of intrinsic
magnetic moment and external magnetic field. The effect from the
couplings among the orbit and the spin to the electric field is
given in the third term.

Previous discussion of spin Hall effect was given in the case of
absence of the magnetic field $\mathbf{B}$. In general, to extend
the cases for the ferromagnet or the system under the external
magnetic field, the magnetic field is remained in the follows and
demonstrated the effect of magnetic field on the spin Hall effect.
Considering an external magnetic field along the direction of the
spin, one state of the spin polarization is left and all spin
transport processes in the presence of both the electric field
$E^{m}$ and the magnetic field $B^{l}$ are shown in Fig. 1(b). The
corresponding the spin motive force and the spin Hall coefficient
can be obtained. It is worth to point out that the previous spin
current does not contain the contribution of spin torque dipole
\cite{29}. When the torque density is written in the form of a
divergence of a torque dipole $T^{l}=-\nabla ^{k}P^{kl}_{T}$, where
$P^{kl}_{T}=\int_{v}T^{l}dx^{k}$ is integrable, the spin current is
found
\begin{equation}
J^{kl}_{s}=j^{kl}_{s}+P^{kl}_{T},
\end{equation}
which includes the traditional current and a correction of the spin
torque dipole. Eq. (4) can be written as a response equation
$J_{s}^{kl}= \sigma_{sc} \varepsilon^{lkm}E^{m}$ in which
$\sigma_{sc}$ is the spin Hall coefficient. Obviously, the spin
current $J^{kl}_{s}$ is vertical to the direction of the spin
$s^{l}$ and the electric field $E^{m}$. $E^{m}$, $s^{l}$, and
$J^{kl}_{s}$ satisfy the right-hand rule, as shown in Fig. 1(a).

Now the spin continuity equation (1) can be written as
\begin{equation}
\frac{\partial}{\partial t}\rho^{l}_{s}+\nabla^{k}J^{kl}_{s}=0.
\end{equation}
It implies that the spin current has a natural conjugate spin force.
Therefore, the Onsager relation $\sigma_{sc}^{mk}=
-\sigma_{cs}^{km}$ can be established under the time reversal
symmetry to link the spin transport with other transport phenomena,
such as the charge transport, where $\sigma_{sc}^{mk}$ and
$\sigma_{cs}^{km}$ are the spin-charge and charge-spin conductivity
tensors.

\section{Spin Hall coefficient and spin motive force}
We consider the divergence of the spin torque dipole as a product of
a
electric field and a coefficient $\chi^{lm}(\mathbf{q})$, $-iq^{k}%
P_{T}^{kl}(\mathbf{q})=\chi^{lm}(\mathbf{q})E^{m}(\mathbf{q})$, with
$\mathbf{q}$ being a finite wave vector. The more explicit form of
the coefficient can be represented as follow
\begin{eqnarray}
\fl \chi^{lm}= -\frac{\hbar}{2}\varepsilon^{lm^{^{\prime}}m}q
^{m^{^{\prime}}}\frac{\sigma_{e}}{e}+\frac{i\hbar e}{2mc}\varphi^{+}(\mathbf{q})%
[(\sigma^{l}\sigma\cdot\mathbf{B} -B^{l})/E^{m}]\varphi(\mathbf{q})
\nonumber\\
-\frac{\hbar e}{4m^{2}c^{2}}\varphi^{+}(\mathbf{q})(i\hbar q^{m^{^{\prime}}}%
\sigma^{m}\sigma^{n^{^{\prime}}}\varepsilon^{lm^{^{\prime}}n^{^{\prime}}%
}+2\sigma^{l}\pi^{m})\varphi(\mathbf{q}),
\end{eqnarray}
where $\sigma_{e}$ is the electric conductivity. The spin Hall
coefficient $\sigma_{sc}$ corresponding to our new spin current
$J_{s}^{kl}$ can be written as
\begin{equation}
\sigma_{sc}=\sigma_{SH}^{0}+\sigma_{SH}^{T},
\end{equation}
where $\sigma_{SH}^{0}$ is the conventional spin Hall conductivity
\cite{1,30}, corresponding to the traditional spin current,
$\sigma_{SH}^{T}$ is the contribution of the spin torque dipole
$P_{T}^{kl}$, and $\sigma
_{SH}^{T}=Re\{i\partial\chi^{lm}(\mathbf{q})/\partial
q^{k}\}_{\mathbf{q}=0}$. In some semiconductors with disorder the
spin Hall coefficient is extremely different from the conventional
one. We can evaluate the spin Hall coefficient $\sigma_{SH}^{T}$ in
the GaAs sample as follows: at
room temperature, the carrier density of GaAs is $n\sim\mathrm{10^{17}%
}\mathrm{cm}^{\mathrm{-3}}$, the mobility of carriers is $\mu\sim
\mathrm{350cm^{2}}/\mathrm{Vs}$, the conventional spin Hall
coefficient is
$\sigma_{SH}^{0}\sim16\Omega^{-1}\mathrm{cm^{-1}}$, $\sigma_{SH}^{T}%
\sim\mathrm{5.6}\Omega^{\mathrm{-1}}\mathrm{cm}^{\mathrm{-1}}$. For
lower carrier density case, $n\sim\mathrm{10^{16}cm^{-3}}$, $\mu\sim
\mathrm{400cm^{2}/Vs}$,
$\sigma_{SH}^{0}\sim7.3\Omega^{-1}\mathrm{cm^{-1}}$,
$\sigma_{SH}^{T}$ is estimated as $\sigma_{SH}^{T}\sim\mathrm{0.64}%
\Omega^{\mathrm{-1}}\mathrm{cm}^{-1}$. As a kind of correction,
$\sigma _{SH}^{T}$ is one order smaller than the conventional spin
Hall coefficient $\sigma_{SH}^{0}$. The general spin Hall
coefficient $\sigma_{sc}$ should include the conventional one
$\sigma_{SH}^{0}$ and the correction $\sigma _{SH}^{T}$.

Now the Onsager relation and spin Hall coefficient have been found.
The
so-called spin force $\mathbf{F}_{s}$ can be calculated as $\mathbf{F}%
_{s}=(\mathbf{J}_{c}-\sigma_{cc}\mathbf{E})/\sigma_{cs}$, where
$\sigma_{cc}$ is the charge-charge conductivity tensor, and
$\mathbf{J}_{c}$ is charge current \cite{29}. Particularly, in Ref.
\cite{30}, the spin force has a simple form as
$F_{s}^{m}=J^{k}_{c}/\sigma_{cs}^{km}$ in the two-dimensional
electron gas. From Onsager relation,
$\sigma^{km}_{cs}=-\sigma^{mk}_{sc}$, the charge-spin tensor
$\sigma_{cs}^{km}$ can be obtained, and the intrinsic Hall current
$J^{k}_{c}$ in the $k$ direction can be detected by experiments.
However, the spin force can not be interpreted as a motive force of
electron like the Lorentz force in Hall effect, and it has the same
direction with the electric field $E^{m}$.

To interpret the spin Hall effect, we try to find a spin motive
force $f^{k}$ which has an analogy to the Lorentz force in the Hall
effect. Here the spin motive force is vertical to the direction of
the electric field and the spin, i.e., $E^{m}$, $s^{l}$ and $f^{k}$
satisfy the right-hand rule, as shown in Fig. 1(a). The discussion
is based on the spin torque. The torque density
$T^{l}$ can be written as the form $T^{l}=\varepsilon^{lmk}r^{m}f^{k}%
=\chi^{lm}E^{m}$. After calculation, we obtain $f^{k}$ as
\begin{equation}
f^{k}=\sigma_{f}^{1}E^{m}+\sigma_{f}^{2}\chi^{lm},
\end{equation}
where the spin motive force coefficients $\sigma_{f}^{1}$ and
$\sigma_{f}^{2}$ are expressed as
\begin{equation}
\sigma_{f}^{1} =\frac{1}{2}Re\{\varepsilon^{lmk}\nabla^{m}\chi^{lm}%
(\mathbf{r})\}
\end{equation}
and
\begin{equation}
\sigma_{f}^{2}
=\frac{1}{2}Re\{\varepsilon^{lmk}\nabla^{m}E^{m}(\mathbf{r})\}.
\end{equation}
In the case of the electric field being constant, $\sigma_{f}^{2}$
is zero. Here we have obtained the evident formula $\chi^{lm}$, and
the electric field $E^{m}$ can be detected in experiments. Thus the
spin motive force $f^{k}$ is found. Assuming the mobility of the
carriers in the GaAs sample with disorder
is $\mu\sim\mathrm{10^{3}cm^{2}/Vs}$ and the electric field is $\mathbf{E}%
\sim\mathrm{10mV/\mu m}$, we find the order-of-magnitude of the spin
motive force $f^{k}\sim\mathrm{10^{-20}eV/\mu m}$. Obviously, this
is an extremely weak quantity.

\section{Application in the two-dimensional electron gas}
We will discuss the properties of the spin motive force in the
two-dimensional electron gas. The Dirac Hamiltonian of relativistic
electron is $H=c \alpha \cdot\mathbf{P}+\beta mc^{2}$. Using the F-W
transformation, the nonrelativistic limit of the Dirac Hamiltonian
is
$H=\beta(mc^{2}+\pi^{2}/2m-\pi^{4}/8m^{3}%
c^{2})+e\phi-\left(  \hbar e/2mc\right) \beta\sigma\cdot
\mathbf{B}-\left(  \hbar^{2}e/8m^{2}c^{2}\right)  \nabla\cdot\mathbf{E}%
-i\left(  \hbar^{2}e/8m^{2}c^{2}\right) \sigma\cdot(\nabla
\times\mathbf{E})-\left(  \hbar e/4m^{2}c^{2}\right) \sigma\cdot
(\mathbf{E}\times\mathbf{P})$, where $\phi$ is the electric
potential \cite{32}. In the two-dimensional electron gas,
$\mathbf{E}=(0,0,E^{m})$,
$\sigma=(\sigma^{l},\sigma^{k},\sigma^{m})$, $\mathbf{P}%
=(P^{l},P^{k},0)$, and $\mathbf{B}=0$, the nonrelativistic
Hamiltonian can be written as
$H=\mathbf{P}^{2}/2m-\lambda(P^{k}\sigma^{l}-P^{l}\sigma^{k})$, this
is the Rashba Hamiltonian, where the coupling parameter
$\lambda=\left( \hbar e/4m^{2}c^{2}\right)  E^{m}$ \cite{35}.

In the two-dimensional electron gas, the formula $\chi^{lm}$ has a
simple
form, $\chi^{lm}=i\left(  \hbar/2e\right)  \varepsilon^{lm^{^{\prime}}m}%
\nabla^{m^{^{\prime}}}\sigma_{e}-\left(  \hbar e/4m^{2}c^{2}\right)
\varphi^{+}(\hbar\nabla^{m^{\prime}}\sigma^{m}\sigma^{n^{\prime}}%
\varepsilon^{lm^{\prime}n^{\prime}}+2\sigma^{l}\pi^{m})\varphi$.
Thus the spin
motive force can be represented as $f^{k}=\left( \varepsilon^{lkm} \hbar e/8m^{2}%
c^{2}\right)  \nabla^{m}[\varphi^{+}(\hbar\nabla^{m^{\prime}%
}\sigma^{m}\sigma^{n^{\prime}}\varepsilon^{lm^{\prime}n^{\prime}}+2\sigma
^{l}\pi^{m})\varphi] E^{m}$. The spin motive force $f^{k}$ is
nonzero, and it induces the spin current, so the spin Hall effect
can be observed in experiments in the two-dimensional electron gas.
In this case, $f^{k}$ should be vertical to the spin $s^{l}$ and
electric field $E^{m}$, as shown in Fig. 1(c). In Ref. \cite{33},
the author introduced a spin transverse force which is perpendicular
to the spin current. On the contrary, our spin motive force is
parallel to the spin current. So it can be used to better understand
the mechanism of the spin Hall effect.

In conclusion, we induce the spin continuity equation from the
angular momentum conservation law with spin-orbit coupling. Our
results naturally include a correction to the traditional spin
current. The correction could be considered as a spin torque dipole,
so there is a conjugate force linking the spin current, and the
Onsager relation can be established. A perfect spin Hall coefficient
corresponding to the new spin current is conformed. Furthermore, the
magnitude of the spin Hall coefficient is evaluated. From the
explicit spin torque, we introduce a spin motive force having the
same direction with the spin current to better understand the spin
Hall effect. We find a novel right-hand rule among the electric
field, the spin and spin current (or spin motive force) in
spintronics.

We are grateful to Z. S. Ma and Y. G. Yao for helpful discussions.
This work was supported by NSF of China under grant 10347001,
90403034, 90406017, 60525417, 10665003, and by NKBRSF of China under
2005CB724508 and 2006CB921400.


\section*{References}
\begin{thebibliography}{10}
\bibitem{22} $\check{\mathrm{Z}}$uti$\acute{\mathrm{c}}$ I, Fabian J and Sarma S D 2004
{\it Rev. Mod. Phys.} {\bf 76} 323

\bibitem{23} Wolf S A, Awschalom D D, Buhrman R A, Daughton J M,
von Molnar S, Roukes M L, Chtchelkanova A Y, and Treger D M 2001
{\it Science} {\bf 294} 1488

\bibitem{24} Prinz G A 1998 {\it Science} {\bf 282} 1660

\bibitem{1}  Murakami S, Nagaosa N and Zhang s c 2003 {\it Science} {\bf 301} 1348

\bibitem{4}  Sinova J, Culcer D, Niu Q, Sinitsyn N A, Jungwirth T,
and MacDonald A H 2004 {\it Phys. Rev. Lett.} {\bf92} 126603

\bibitem{29} Shi J, Zhang P, Xiao D, and Niu Q 2006 {\it Phys. Rev. Lett.} {\bf 96} 076604

\bibitem{25} Sun Q F and Xie X C 2005 {\it Phys. Rev. B} {\bf72} 245305

\bibitem{26} Wang Y, Xia K, Su Z B, and Ma Z 2006 {\it Phys. Rev. Lett.} {\bf 96} 066601

\bibitem{27} Zhang S and Yang Z 2005 {\it Phys. Rev. Lett.} {\bf 94} 066602

\bibitem{28} Culcer D, Sinova J, Sinitsyn N A, Jungwirth T,
MacDonald A H and Niu Q 2004 {\it Phys. Rev. Lett.} {\bf93} 046602

\bibitem{36} Shen R, Chen Y, Wang Z D, Xing D Y 2006 {\it Phys. Rev. B} {\bf 74} 125313

\bibitem{30} Zhang P and Niu Q cond-mat/0406436

\bibitem{2} Dyakonov M I and Perel V I 1971 {\it Sov. Phys. JETP} {\bf 13} 467

\bibitem{3} Hirsch J E 1999 {\it Phys. Rev. Lett.} {\bf 83} 1834

\bibitem{5} Rashba E I 2004 {\it Phys. Rev. B} {\bf 70} 161201

\bibitem{6} Hu J P, Bernevig B A, and Wu C J 2003 {\it Int. J. Mod. Phys. B} {\bf 17} 5991

\bibitem{7} Sinitsyn N A, Hankiewicz E M, Teizer W and Sinova J 2004 {\it Phys. Rev. B} {\bf 70} 081312

\bibitem{8} Bernevig B A 2005 {\it Phys. Rev. B} {\bf 71} 073201

\bibitem{9} Shen S Q, Ma M, Xie X C, and Zhang F C 2004 {\it Phys. Rev. Lett.} {\bf 92} 256603

\bibitem{10} Guo G Y, Yao Y and Niu Q 2005 {\it Phys. Rev. Lett.} {\bf 94} 226601

\bibitem{11} Inoue J I, Bauer G E W and Molenkamp L W 2004 {\it Phys. Rev. B} {\bf 70} 041303(R)

\bibitem{12} Mishchenko E G, Shytov A V, and Halperin B I 2004 {\it Phys. Rev. Lett.} {\bf 93} 226602

\bibitem{13} Dimitrova O V 2005 {\it Phys. Rev. B} {\bf 71} 245327

\bibitem{14} Chalaev O and Loss D 2005 {\it Phys. Rev. B} {\bf 71} 245318

\bibitem{15} Bernevig B A and Zhang S C 2005 {\it Phys. Rev. Lett.} {\bf 95} 016801

\bibitem{16} Rashba E I 2003 {\it Phys. Rev. B} {\bf 68} 241315(R)

\bibitem{17} Jin P Q, Li Y Q and Zhang F C 2006 {\it J. Phys. A} {\bf 39} 7115

\bibitem{18} Wang J, Wang B G, Ren W and Guo H cond-mat/0507159

\bibitem{19} Kato Y K, Myers R C, Gossard A C and Awschalom D D 2004 {\it Science} {\bf 306} 1910

\bibitem{20} Wunderlich J, Kaestner B, Sinova J and Jungwirth T 2005 {\it Phys. Rev. Lett.} {\bf 94} 047204

\bibitem{21} Liu S Y and Lei X L 2005 {\it Phys. Rev. B} {\bf 72} 155314

\bibitem{34} Sugimoto N, Onoda S, Murakami S and Nagaosa N cond-mat/0503475

\bibitem{31} Foldy L L and Wouthuysen S A 1950 {\it Phys. Rev.} {\bf 78} 29

\bibitem{32} Bjorken J D and Drell S D 1964 {\it Relativistic Quantum
Mechanics} (New York: Mc Graw-Hill)

\bibitem{35} Bychkov Y A and Rashba E I 1984 {\it J. Phys. C} {\bf 17} 6039

\bibitem{33} Shen S. Q. 2005 {\it Phys. Rev. Lett.} {\bf95}, 187203
\end{thebibliography}
\end{document}